\documentclass[aps,twocolumn,superscriptaddress,showpacs,amsmath,amssymb]{revtex4-1}

\usepackage{graphicx}
\usepackage{dcolumn}
\usepackage{bm}
\usepackage{color}
\begin{document}

\title{Low-temperature thermal conductivity and magnetic transitions of the kagom\'{e}-staircase compound Ni$_3$V$_2$O$_8$}

\author{Z. Y. Zhao}
\affiliation{Department of Physics, Hefei National Laboratory for Physical Sciences at Microscale, and Key Laboratory of Strongly-Coupled Quantum Matter Physics (CAS), University of Science and Technology of China, Hefei, Anhui 230026, People's Republic of China}
\affiliation{State Key Laboratory of Structural Chemistry, Fujian Institute of Research on the Structure of Matter, Chinese Academy of Sciences, Fuzhou, Fujian 350002, People's Republic of China}

\author{Q. J. Li}
\affiliation{School of Physics and Materials Science, Anhui University, Hefei, Anhui 230601, People's Republic of China}

\author{X. G. Liu}
\affiliation{Hefei National Laboratory for Physical Sciences at Microscale, University of Science and Technology of China, Hefei, Anhui 230026, People's Republic of China}

\author{X. Rao}
\affiliation{Department of Physics, Hefei National Laboratory for Physical Sciences at Microscale, and Key Laboratory of Strongly-Coupled Quantum Matter Physics (CAS), University of Science and Technology of China, Hefei, Anhui 230026, People's Republic of China}

\author{H. L. Che}
\affiliation{Department of Physics, Hefei National Laboratory for Physical Sciences at Microscale, and Key Laboratory of Strongly-Coupled Quantum Matter Physics (CAS), University of Science and Technology of China, Hefei, Anhui 230026, People's Republic of China}

\author{L. G. Chu}
\affiliation{Department of Physics, Hefei National Laboratory for Physical Sciences at Microscale, and Key Laboratory of Strongly-Coupled Quantum Matter Physics (CAS), University of Science and Technology of China, Hefei, Anhui 230026, People's Republic of China}

\author{Z. Z. He}
\affiliation{State Key Laboratory of Structural Chemistry, Fujian Institute of Research on the Structure of Matter, Chinese Academy of Sciences, Fuzhou, Fujian 350002, People's Republic of China}

\author{X. Zhao}
\email{xiazhao@ustc.edu.cn}
\affiliation{School of Physical Sciences, University of Science and Technology of China, Hefei, Anhui 230026, People's Republic of China}

\author{X. F. Sun}
\email{xfsun@ustc.edu.cn}
\affiliation{Department of Physics, Hefei National Laboratory for Physical Sciences at Microscale, and Key Laboratory of Strongly-Coupled Quantum Matter Physics (CAS), University of Science and Technology of China, Hefei, Anhui 230026, People's Republic of China}

\affiliation{Institute of Physical Science and Information Technology, Anhui University, Hefei, Anhui 230601, People's Republic of China}

\affiliation{Collaborative Innovation Center of Advanced Microstructures, Nanjing University, Nanjing, Jiangsu 210093, People's Republic of China}

\date{\today}

\begin{abstract}

Kagom\'{e}-staircase compound Ni$_3$V$_2$O$_8$ is an attractive multiferroic material exhibiting rich phase diagrams. However, the magnetic properties and magnetic transitions have been studied only above 1.3 K. In this work, we study the thermal conductivity ($\kappa$) of Ni$_3$V$_2$O$_8$ single crystals at low temperatures down to 0.3 K and in magnetic fields up to 14 T. In zero field, the magnetic transitions from the low-temperature incommensurate (LTI) phase to the commensurate phase (C) and then to a second commensurate phase (C') yield anomalies in $\kappa(T)$ curves at $T\rm_{LC}$ = 3.7 K and $T\rm_{CC'}$ = 2.0 K, respectively, which indicates a significant phonon scattering by the critical spin fluctuations. When the field is applied along the $a$ axis, the field dependence of $\kappa$ displays four anomalies associated with different magnetic transitions and reveals an undetected magnetic state at subKelvin temperatures. In addition, the $\kappa(B)$ curves are found to depend not only on the history but also on the magnitude of applying field. When the field is applied along the $b$ axis, a high-field phase locating above the LTI and high-temperature incommensurate (HTI) phases is revealed.

\end{abstract}

\pacs{66.70.-f, 75.47.-m, 75.50.-y}

\maketitle

\section{Introduction}

Multiferroicity refers to a coupling effect between the magnetization ($M$) and electric polarization ($P$). The mutual control of the magnetism and electricity has received considerable attention due to the potential applications \cite{Multiferroic-1, Multiferroic-2, Multiferroic-3, Multiferroic-4, Multiferroic-5}. In comparison to the type-I multiferroics, in which $M$ and $P$ are originated from different sublattices, the magnetoelectric coupling of type-II multiferroics can be much stronger and therefore is of great interest to condensed matter physicists. Ni$_3$V$_2$O$_8$ with a kagom\'{e}-staircase lattice is an outstanding improper multiferroics exhibiting rich phase diagrams. The deviation from the ideal kagom\'{e} geometry results in two inequivalent Ni$^{2+}$ sites. ``Spine" (Ni$_s$) sites form chains running along the $a$ axis and are connected by the ``cross-tie" (Ni$_c$) spins in the $c$ direction \cite{Structure}, see the inset to Fig. \ref{kappa-0T}. In zero magnetic field, Ni$_3$V$_2$O$_8$ experiences a cascade of antiferromagnetic transitions when the temperature is reduced \cite{PRL-93, PRB-74}. Below $T\rm_{PH}$ = 9.1 K, Ni$_s$ spins firstly develop an incommensurate order (HTI) with a longitudinally modulated spin structure along the $a$ axis. Upon cooling the system enters into a second incommensurate phase (LTI) at $T\rm_{HL}$ = 6.3 K, and both Ni$_s$ and Ni$_c$ spins are spirally ordered in the $ab$ plane. A spin-order induced $P$ along the $b$ axis is accordingly introduced due to the breaking of the spatial inversion symmetry \cite{PRL-95}. With further lowering temperature, two commensurate phases C and C' set in at $T\rm_{LC}$ = 3.9 K and $T\rm_{CC'}$ = 2.2 K, respectively, and the staggered moment points along the $a$ axis with a weak ferromagnetic component along the $c$ axis. The transitions at $T\rm_{PH}$, $T\rm_{HL}$, and $T\rm_{CC'}$ are the second order, while the transition at $T\rm_{LC}$ is the first order. Although there was a pronounced peak in the specific heat at $T\rm_{CC'}$, the difference between C and C' phases remained unclear until a recent elastic neutron scattering study \cite{C'}. The ground state was discussed to be a mixture of incommensurate Ni$_c$ and commensurate Ni$_s$ spin orderings, in contrast to the solely commensurate magnetism as previously proposed.

The application of magnetic field can give rise to multiple magnetic transitions \cite{PRL-93}. However, as far as we know, the $B-T$ phase diagrams for fields along the principle crystallographic directions have been explored only at temperatures above 1.3 K \cite{High_field_parallel_a-1, High_field_parallel_b-2}. The magnetism and magnetic transitions at subKelvin temperatures and in high magnetic fields have not been probed. In particular, the argument about the high-field phases above 2 K for $B \parallel b$ is still an open question \cite{High_field_parallel_b-1}. Heat transport has been proved to be a useful probe to explore the low-temperature magnetic transitions in multiferroic materials \cite{GdFeO3, DyFeO3, TmMnO3, HoMnO3, ErMnO3, CuFeO2}, in the regard that it is much easier to carry out at very low temperatures and in high fields than some other measurements. In this work, the temperature and field dependencies of the thermal conductivity ($\kappa$) is measured down to 0.3 K and up to 14 T to study the magnetic transitions and construct the anisotropic phase diagrams.

\section{Experiments}

High-quality Ni$_3$V$_2$O$_8$ single crystals were grown using flux method \cite{Crystal_growth}. Long-bar shaped samples were cut along the principle crystallographic directions. Thermal conductivity was measured using a ``one heater, two thermometers" technique in a $^3$He refrigerator and a 14 T magnet at temperature regime of 0.3--8 K and using a Chromel-Constantan thermocouple in a $^4$He pulse-tube refrigerator in 0 T above 4 K \cite{DTN, CuFeO2, Co3V2O8}. In these measurements, heat current was always applied along the length of the sample with the magnetic field applied in a parallel or perpendicular configuration.

\section{Results and discussions}

\subsection{Thermal conductivity in zero field}

\begin{figure}
\includegraphics[clip,width=7.5cm]{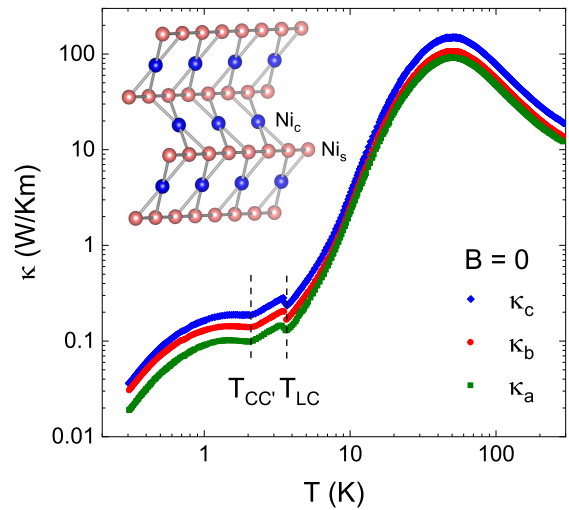}
\caption{(Color online) Zero-field $\kappa$ measured along the principle crystallographic directions. $T\rm_{LC}$ and $T\rm_{CC'}$ are the critical temperatures from the incommensurate LTI phase to the commensurate C phase and then to the second commensurate C' phase. The inset is the kagom\'{e}-staircase geometry with inequivalent spine (pink) and cross-tie (blue) Ni$^{2+}$ spins.}
\label{kappa-0T}
\end{figure}

Figure \ref{kappa-0T} shows the temperature dependencies of $\kappa$ measured along the principle crystallographic directions in zero field. $\kappa$ is nearly isotropic for three directions. The high-$T$ $\kappa$ has large magnitude and a maximum is present around 50 K with magnitude about 90 W/Km for $\kappa_a$, 110 W/Km for $\kappa_b$, and 150 W/Km for $\kappa_c$, respectively. It should be noted that the usual phonon peak locates at about 20 K, which means that phonons suffer strong scattering effect other than crystal imperfections even at rather high temperatures. This is likely due to the magnetic scattering effect, which is commonly found in some other complex magnetic materials \cite{BaCo2V2O8, Co3V2O8, CuFeO2}. As temperature is lowered, two anomalies are observed at $T\rm_{LC}$ = 3.7 K and $T\rm_{CC'}$ = 2.0 K, which are associated with the transitions from LTI to C phase and C to C' phase. The minimum feature of the anomalies is indicative of significant phonon scattering by the critical spin fluctuations across the magnetic transitions. In spite of four magnetic transitions at low temperatures, only two of them are observable in $\kappa(T)$. Similar situation was also found in the analogous material Co$_3$V$_2$O$_8$, in which only the transition to the ground state was detected in $\kappa(T)$ although a series of magnetic transitions occurred in zero field \cite{Co3V2O8}.

Specific heat is the unique probe to discover all the four magnetic transitions \cite{PRL-93}. However, magnetic susceptibility and thermal expansion can only ``see" the transition at $T\rm_{LC}$, whereas neutron scattering can further capture the transition at $T\rm_{HL}$ \cite{PRL-93, Pressure}. Experimentally, it is difficult to distinguish the C and C' phases at $T\rm_{CC'}$. Until recently, an elastic neutron scattering study stated definitely that the C' phase was a mixture of incommensurate and commensurate orderings \cite{C'}. The incommensurate part was related to Ni$_c$ spins which ordered in two cycloids within $bc$ plane along the $c$ direction, and the commensurate part corresponded to Ni$_s$ spins which were aligned antiferromagnetically along the $a$ axis.

\begin{figure}
\includegraphics[clip,width=7.5cm]{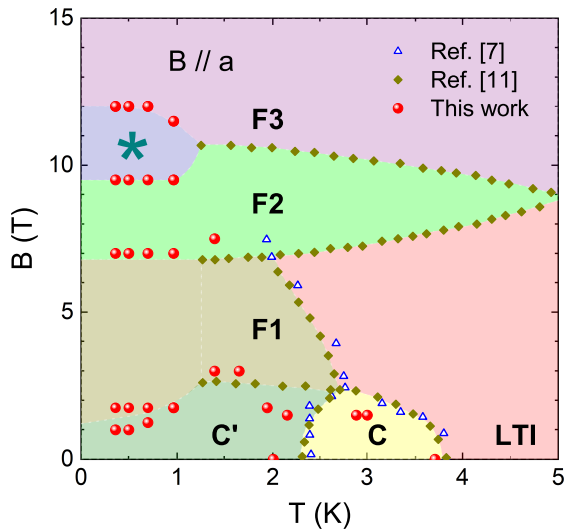}
\caption{(Color online) $B-T$ phase diagram for $B \parallel a$. Filled circles are the critical temperatures and magnetic fields determined in this work. Data indicated by the open triangles and filled diamonds are reproduced from Ref. [\onlinecite{PRL-93}] and Ref. [\onlinecite{High_field_parallel_a-1}] for comparison. ``F1", ``F2" and ``F3" are the unique high-field phases explored by pulsed-field magnetization. ``LTI" represents the low-temperature incommensurate phase, ``C" and ``C'" denote two different commensurate states. The area at subKelvin temperatures, marked by star, is discovered via $\kappa(B)$ data of this work.}
\label{Diagram_Ha}
\end{figure}

Considering that the magnetic ordering of C' phase is partially incommensurate \cite{C'}, the observation of these two transitions in $\kappa(T)$ can be probably attributed to the sensitivity of $\kappa$ to the change of the commensurability of the spin density wave vector, that is, from incommensurate state to commensurate state across $T\rm_{LC}$ and then to incommensurate state across $T\rm_{CC'}$. In the multiferroic TbMnO$_3$, the commensurability change of the spin or electric dipole order has already been reported to influence $\kappa$ \cite{TbMnO3}. The slight change of the incommensurate wave vector from HTI to LTI phase modulates weakly the dispersive spectra of magnetic excitations and hence influences hardly on $\kappa(T)$ at $T\rm_{HL}$ \cite{PRB-74}. The fact that the transition at $T\rm_{CC'}$ is broader than $T\rm_{LC}$ also supports this assumption considering the partial incommensurate ordering of C' phase. It should be noted that the zero-field $\kappa_b$ had been previously reported, in which not only the transition at $T\rm_{CC'}$ was absent but also the anomaly at $T\rm_{LC}$ was very weak \cite{kappa}. The larger magnitude of $\kappa(T)$ and obvious anomalies in the present data indicate high quality of our single crystals.

\subsection{Thermal conductivity in $B \parallel a$}

When applying field along the $a$ direction, the resultant $B-T$ phase diagram is rather complicated. Besides the four ordered phases in low fields, another five distinct phases named from F1 to F5 were discovered by the high-field magnetization up to 30 T at 1.3 K \cite{High_field_parallel_a-1}. Particularly, an apparent 1/2 magnetization plateau was observed in the F3 phase. Electric polarization measurement further confirmed that F1, F2, and F4 phases are ferroelectric (FE) \cite{High_field_parallel_a-2}. The phase diagram below 5 K and up to 15 T is reproduced in Fig. \ref{Diagram_Ha}.

\begin{figure}
\includegraphics[clip,width=7.0cm]{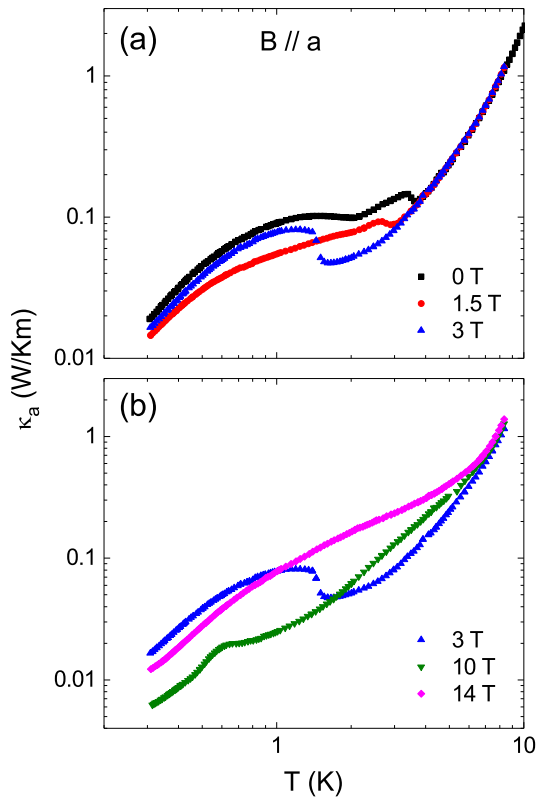}
\caption{(Color online) Temperature dependencies of $\kappa_a$ in $B \parallel a$ below 3 T (a) and above 3 T (b).}
\label{kaT_Ha}
\end{figure}

The $\kappa$ as a function of temperature in different fields is displayed in Fig. \ref{kaT_Ha}. The application of $a$-axis magnetic field induces complex changes on $\kappa(T)$. In 1.5 T, the $\kappa$ in the incommensurate state (C and C' phases) is strongly weakened, and $T\rm_{LC}$ is reduced to 2.9 K while the anomaly at $T\rm_{CC'}$ is almost smeared out along with a very weak slope change at 2.2 K. Both transitions are completely wiped out in 3 T and instead a step-like enhancement is emerged below 1.6 K. As field is further increased, the subKelvin-temperature $\kappa$ is significantly suppressed in 10 T and a hump is present around 0.6 K. In 14 T, the $\kappa$ is somewhat recovered compared to the lower-field data and shows a rather smooth temperature dependence. Note that at temperature above 1 K, the 14 T data become even much larger than the zero-field data, which indicates the suppression of phonon scattering in high field.

\begin{figure}
\includegraphics[clip,width=8.5cm]{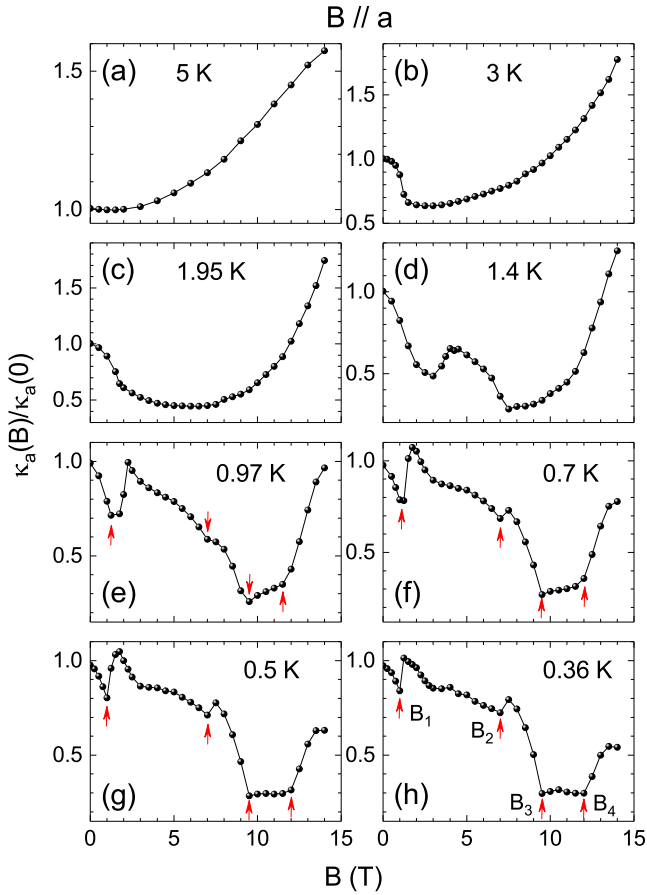}
\caption{(Color online) Magnetic field dependencies of $\kappa_a$ measured in $B \parallel a$ at various temperatures. Data are collected from 14 to 0 T after two steps: (i) the crystal is firstly cooled down to the target temperature in zero field; and then (ii) the magnetic field is swept from 0 to 14 T. Below 1 K, four magnetic transitions labeled as $B_1$ to $B_4$ are indicated by the arrows.}
\label{kaHa}
\end{figure}

Figure \ref{kaHa} shows the field dependence of $\kappa$ measured from 14 to 0 T ($\kappa_\downarrow$) after cooling in zero field. At 5 K, $\kappa$ is gradually enhanced with increasing field. Upon cooling, a step-like decrease appears below 2 T (Figs. \ref{kaHa}(b) and \ref{kaHa}(c)). At 1.4 K, two minimums locating at 3 T and 7.5 T are developed and the magnitude in 14 T is very close to the zero-field value (Fig. \ref{kaHa}(d)). Below 1 K, the $\kappa$ exhibits complicated field dependence as shown in Figs. \ref{kaHa}(e-h). At 0.97 K, four successive magnetic transitions are observed at $B_1$ = 1.25 T, $B_2$ = 7 T, $B_3$ = 9.5 T, and $B_4$ = 11.5 T. With further lowering temperature, $B_2$ and $B_3$ are almost unchanged, while $B_1$ is decreased and $B_4$ is increased. At 0.36 K, the anomaly at $B_2$ evolves into a clear dip and a plateau is formed between $B_3$ = 9.5 T and $B_4$ = 12 T. Here $B_1$ and $B_2$ are defined as the local minimum positions, while $B_3$ and $B_4$ are defined as the starting and ending points of the plateau.

\begin{figure}
\includegraphics[clip,width=8.5cm]{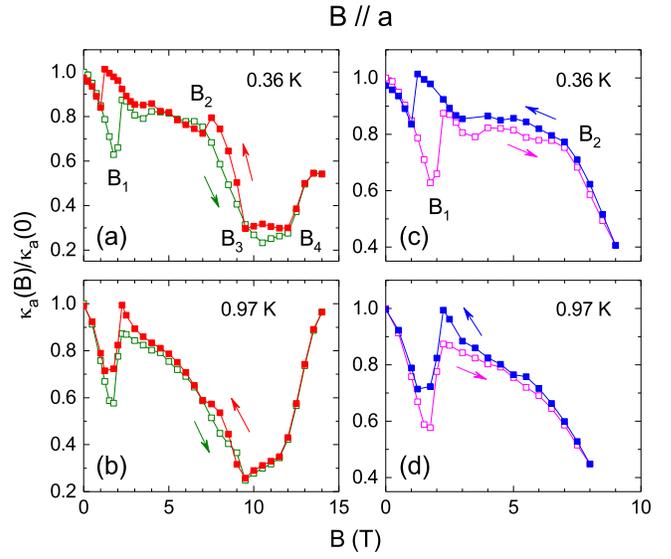}
\caption{(Color online) Magnetic field dependencies of $\kappa_a$ at 0.36 K and 0.97 K in $B \parallel a$. Data are collected in different ways after cooling in zero field: (a-b) ramping field from 0 to 14 T (open square) and then from 14 to 0 T (solid square); (c-d) ramping field from 0 to 9 T at 0.36 K (or 8 T at 0.97 K) (open square) and then down to 0 T (solid square).}
\label{kaHa_hysteresis}
\end{figure}

The most impressive character is the irreversibility related to the history of applying magnetic field. Figures \ref{kaHa_hysteresis}(a) and \ref{kaHa_hysteresis}(b) show the representative $\kappa(B)$ results measured at 0.36 and 0.97 K after cooling in zero field. Here the data collected from 0 to 14 T are named as $\kappa_\uparrow$. The different $B_1$ in $\kappa_\uparrow$ and $\kappa_\downarrow$ result in a very large hysteresis, indicating a first-order transition. Great attention should be paid that all the four magnetic transitions are clearly observable in $\kappa_\downarrow$, while the $\kappa_\uparrow$ do not display clear anomalies at $B_2$ and $B_3$. When the sample is warmed up, the transition field $B_1$ in $\kappa_\uparrow$ is robust but it is gradually increased in $\kappa_\downarrow$, leading to a gradually narrowed hysteresis loop. The dip at $B_2$ in $\kappa_\downarrow$ is also weakened and disappeared above 1 K. In addition, $B_3$ and $B_4$ are gradually merged together, and above 1.4 K there is no difference between $\kappa_\uparrow$ and $\kappa_\downarrow$.

The critical fields and temperatures determined from $\kappa$ are also summarized in Fig. \ref{Diagram_Ha}. The obtained $T\rm_{LC}$ and $T\rm_{CC'}$ are slightly lower than those determined from the specific heat measurements \cite{PRL-93}. The observed anomalies at $B_1$ and $B_2$ are likely related to the magnetic transitions from C' to F1 phase and F1 to F2 phase, respectively. The observed first-order transition from C' to F1 phase is consistent with the magnetization measurement \cite{High_field_parallel_a-1}. It is notable that a new phase sandwiched between F2 and F3 phases is discovered. This phase exists only at subKelvin temperatures and was thus escaped from the investigation in the previous magnetization measurement above 1 K. In view of the separated magnetism of Ni$_c$ and Ni$_s$ spins as suggested by muon-spin relaxation measurement \cite{muSR}, the cascade of field-induced magnetic transitions might be attributed to a step-by-step evolution of the magnetic structures related to Ni$_s$ and Ni$_c$ spins.

From Fig. \ref{kappa-0T}, it is known that the magnetic excitations are effective phonon scatterers and $\kappa(B)$ can exhibit drastic changes across the magnetic phase transitions due to the sudden changes of the population of the magnetic excitations. The first-order transition from C' to F1 phase is therefore responsible for the hysteresis at $B_1$. However, the indistinguishable transitions at $B_2$, $B_3$, and $B_4$ in $\kappa_\uparrow$ are somehow puzzled. Microscopic phonon scatterers like point defects or dislocations are irrelevant since such effect is not influenced by the applied field and becomes less effective at very low temperatures. The peculiar irreversibility of $\kappa(B)$ thus demonstrates some addition channel of phonon scattering that is strongly related to the way of applying field. Similar phenomenon has been already reported in multiferroic GdFeO$_3$ \cite{GdFeO3} and DyFeO$_3$ \cite{DyFeO3}. In GdFeO$_3$, the FE domain walls were considered to be the main source of irreversible phonon scattering. The fact that the field-up $P$ was larger than that in the field-down run implies that there are less FE domains in the field-up process \cite{GdFeO3-2}. The larger $\kappa_\uparrow$ can be attributed to the weaker phonon scattering by the domain walls \cite{GdFeO3}. In DyFeO$_3$, except for the domain wall scattering effect, the presence of a metastable state was also responsible for the hysteresis below 500 mK \cite{DyFeO3}.

It is of special interest to find that $\kappa(B)$ depends not only on the history but also on the magnitude of applying field. Figures \ref{kaHa_hysteresis}(c) and \ref{kaHa_hysteresis}(d) present $\kappa_\uparrow$ and $\kappa_\downarrow$ measured up to 9 T for 0.36 K and 8 T for 0.97 K after cooling in zero field. Note that the $\kappa_\uparrow$ data are identical to the 14-T $\kappa_\uparrow$ below 9 T. It is clearly seen at 0.36 K that the $\kappa_\downarrow$ curve follows the $\kappa_\uparrow$ one and exhibits a large hysteresis at $B_1$. However, only a slope change is found at $B_2$ in $\kappa_\downarrow$, which is rather different from the dip feature as observed in the 14-T $\kappa_\downarrow$.

The above phenomenon seems to have some relationship to that found in the electric polarization experiment. In Ref. \cite{High_field_parallel_a-2}, $P$ started to appear when the system entered into the FE F1 phase. It was found that $P$ was small when applying 5-T pulsed field ($B_1 < B < B_2$), but was promoted significantly in the application of 13-T pulsed field ($B > B_4$). The pinning-depinning of the chiral domain walls was proposed to explain the magnitude effect of applying field on $P$, which might be a possible origin for the peculiar $\kappa(B)$ behavior. FE polarization measured at subKelvin temperatures is therefore called for to examine the emergence of the newly discovered phase and dig out the physics about the unusual hysteresis behavior.

\begin{figure}
\includegraphics[clip,width=8.5cm]{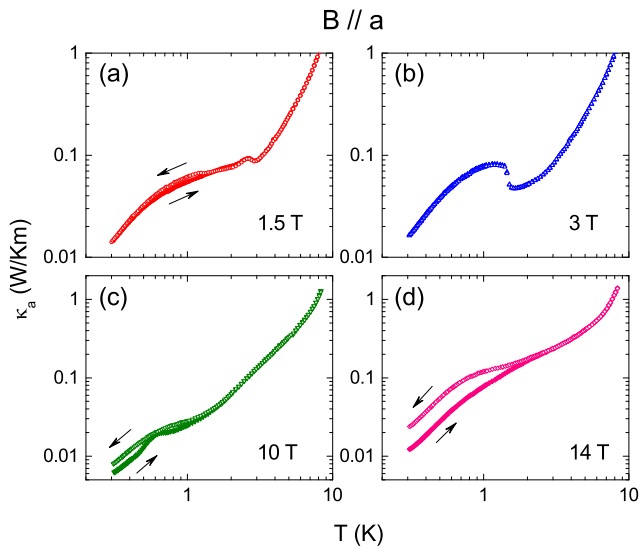}
\caption{(Color online) Temperature dependencies of $\kappa_a$ measured in $B \parallel a$ after cooing in zero field (solid symbol) and selected field (open symbol).}
\label{FC&ZFC}
\end{figure}

It should be pointed out that the $\kappa(B)$ looks like reversible above $B_4$, which is however due to the limitation of the laboratory magnetic field. It is possible that the hysteresis behavior could exist in a wider field window if the field can be applied higher. This expectation can be verified by comparing $\kappa(T)$ with zero-field cooling (ZFC) and field cooling (FC) processes. As shown in Fig. \ref{FC&ZFC}, a weak irreversibility is seen at subKelvin temperature in the magnetic fields exhibiting distinct hysteresis; whereas, the 14-T ZFC and FC $\kappa(T)$ data display a remarkable difference which persists up to 2 K, demonstrating that the 14-T field is
not strong enough to overcome the microscopic origin of the irreversible effect on $\kappa$. This can also be supported by the electric polarization measurement. In Ref. \cite{High_field_parallel_a-2}, $P$ was almost identical above the transition from F2 to F3 phase at 1.6 K if the field was applied only up to 13 T, while a rather obvious hysteresis related to the same transition was observed when the field was pulsed to 30 T.

\subsection{Thermal conductivity in $B \parallel b$}

\begin{figure}
\includegraphics[clip,width=7.5cm]{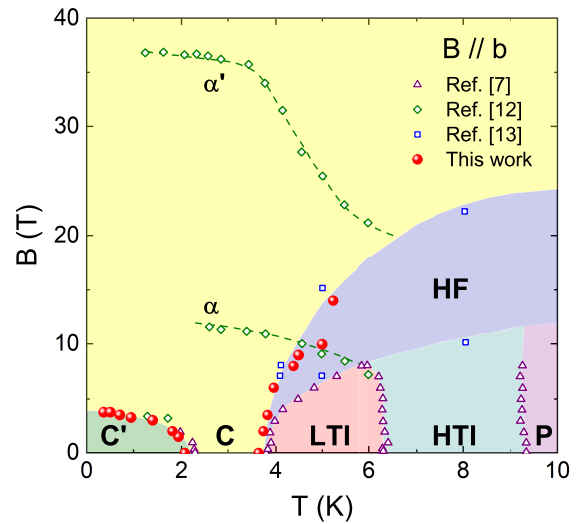}
\caption{(Color online) $B-T$ phase diagram for $B \parallel b$. Filled circles are the critical temperatures and magnetic fields determined in this work. Data indicated by the open triangles, diamonds, and squares are reproduced from Ref. [\onlinecite{PRL-93}], Ref. [\onlinecite{High_field_parallel_b-2}], and Ref. [\onlinecite{High_field_parallel_b-1}] for comparison. ``$\alpha$" and ``$\alpha'$" are the magnetic transitions detected by pulsed-field magnetization. ``HF" is a high-field phase revealed by magneto-optical spectra.``P" is the paramagnetic state, ``HTI" and ``LTI" represent the high- and low-temperature incommensurate phases, ``C" and ``C'" denote two different commensurate states.}
\label{Diagram_Hb}
\end{figure}

$B-T$ phase diagram for field along the $b$ direction is controversial and there are some disagreements about the field-induced magnetic transitions. As shown in Fig. \ref{Diagram_Hb}, a high-field phase (HF) extended above LTI, HTI, and paramagnetic (P) phases was revealed in a magneto-optical spectra study \cite{High_field_parallel_b-1}. However, pulsed-field magnetization indicated another two transitions called $\alpha$ and $\alpha'$ at $T <$ 6 K, and some weak anomalies were also detected in between (not shown) \cite{High_field_parallel_b-2}. Since there are few works about the magnetic transitions under the $b$-axis field, various experiments performed in high fields are demanded to catch on the field-induced magnetism.

\begin{figure}
\includegraphics[clip,width=7.5cm]{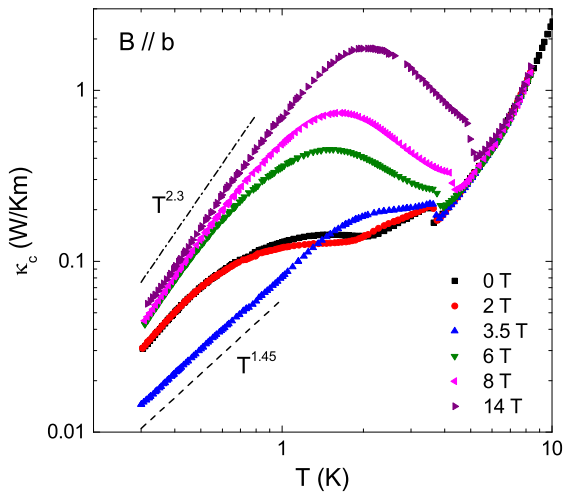}
\caption{(Color online) Temperature dependencies of $\kappa_c$ in $B \parallel b$. At subKelvin temperatures, the $\kappa$ follows a $T^{1.45}$ and $T^{2.3}$ dependence in 3.5 T and 14 T, respectively, as shown by the dashed and dash-dot lines.}
\label{kcT_Hb}
\end{figure}

Temperature dependencies of $\kappa$ measured in various magnetic fields are present in Fig. \ref{kcT_Hb}. $T\rm_{LC}$ and $T\rm_{CC'}$ show opposite trends as the field is increased. $T\rm_{LC}$ is increased with ramping up the field and reaches 5.2 K in 14 T. On the contrary, $T\rm_{CC'}$ is reduced to 1.8 K in 2 T and almost gone in 3.5 T. $\kappa$ is significantly suppressed in 3.5 T, and an approximate $T^{1.45}$ relation is found at subKelvin temperatures. With further increasing fields, the anomaly at $T\rm_{LC}$ becomes sharper and a broad peak is developed below the transition. This is rather distinct from the low-field behavior and probably suggests a different magnetic origin. In 14 T, a $T^{2.3}$ dependence is observed below 1 K. The deviation from the $T^3$ boundary scattering limit \cite{Berman} demonstrates that the microscopic phonon scattering is not negligible. Spin fluctuations originated from the geometrical frustration is possibly the key reason for the persistent dynamics at subKelvin temperatures.

\begin{figure}
\includegraphics[clip,width=7.0cm]{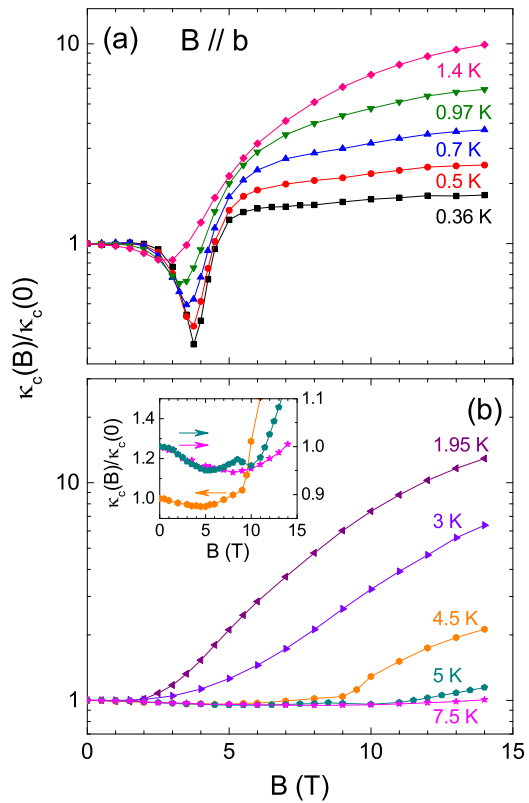}
\caption{(Color online) Magnetic field dependencies of $\kappa_c$ measured in $B \parallel b$ (a) below 1.4 K and (b) above 1.95 K. Inset to panel (b) is the zoom in of the data at 4.5, 5, and 7.5 K.}
\label{kcHb}
\end{figure}

Figure \ref{kcHb} shows the magnetic field dependencies of $\kappa$ at various temperatures. Different from $B \parallel a$, there is no hysteresis observed for $B \parallel b$. At 0.36 K, a dip-like feature is located at 3.75 T with the amplitude suppressed to 30\% of the zero-field value. Upon warming, the dip moves toward to lower fields and the amplitude is gradually weakened. At 3 K, the dip feature vanishes and $\kappa$ is monotonously increased with field. At higher temperatures, some weak features are emerged again. As shown in the inset to Fig. \ref{kcHb}(b), a slope change is found around 9 T at 4.5 K and it is enhanced to 10 T at 5 K. However, the broad and shallow valley appeared at lower fields, which is shifted to higher fields with increasing temperature, is likely resulted from the scattering by paramagnetic moments \cite{GdFeO3,Nd2CuO4,NCCO,Nd3Ga5SiO14}. The ratio of $\kappa(B)/\kappa(0)$ at 14 T is varied nonmonotonously with temperature. The highest ratio is about 13 at 1.95 K, at which temperature the broad peak in $\kappa(T)$ is present. This ratio is very large as compared with other common insulators \cite{BaCo2V2O8,GdFeO3,IPA-CuCl3,NCCO,Nd2CuO4,R2Ti2O7,TmMnO3}, but it is much smaller than Co$_3$V$_2$O$_8$, in which $\kappa(B)/\kappa(0)$ is as high as 100 at 14 T \cite{Co3V2O8}.

As seen in Fig. \ref{Diagram_Hb}, the dip-like feature in $\kappa(B)$ below 2 K corresponds to the transition from C' to C phase. Between $T\rm_{CC'}$ and $T\rm_{LC}$, for example at 3 K, $\kappa(B)$ is monotonously increased up to 14 T without any indication of the $\alpha$-transition. Above $T\rm_{LC}$, the upper boundary of the HF phase is detected through our $\kappa$ measurement. The absent anomaly across the lower boundary probably implies an incommensurate HF phase and $\kappa$ is insensitive to the transition from LTI (or HTI) to HF phase. For clarity, the onset temperature from HF to C phase is still used as $T\rm_{LC}$ in this work. Since there is no hysteresis observed in $\kappa(B)$, the transition from HF state to C state is of second-order nature.

\subsection{Thermal conductivity in $B \parallel c$}

\begin{figure}
\includegraphics[clip,width=7.5cm]{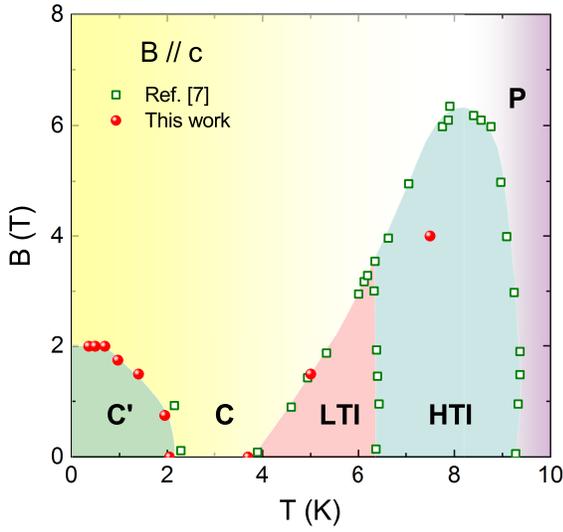}
\caption{(Color online) $B-T$ phase diagram in $B \parallel c$. Filled circles are the critical temperatures and magnetic fields determined in this work. Data indicated by the open squares are reproduced from Ref. [\onlinecite{PRL-93}] for comparison. ``P" is the paramagnetic state, ``HTI" and ``LTI" represent the high- and low-temperature incommensurate phases, ``C" and ``C'" denote two different commensurate states.}
\label{Diagram_Hc}
\end{figure}

\begin{figure}
\includegraphics[clip,width=7.5cm]{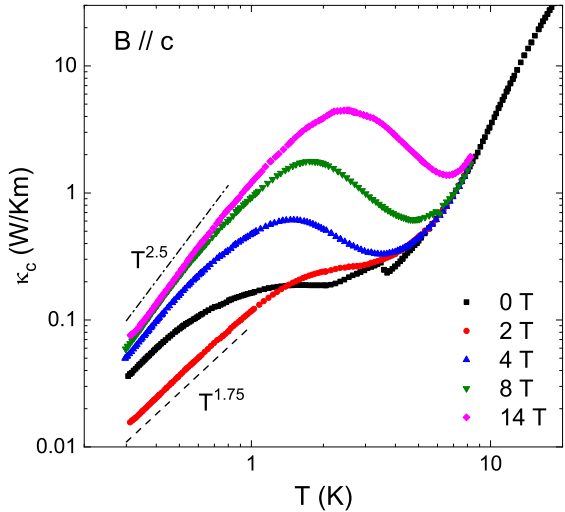}
\caption{(Color online) Temperature dependencies of $\kappa_c$ for $B \parallel c$. At subKelvin temperatures, $\kappa_c$ follows a $T^{1.75}$ and $T^{2.5}$ dependence in 2 T and 14 T, respectively, as shown by the dashed and dash-dot lines.}
\label{kcT_Hc}
\end{figure}

$B-T$ phase diagram for field along the $c$ direction has been reported to be the simplest, as seen in Fig. \ref{Diagram_Hc}. The V-shaped C phase is extended to high fields, and no extra magnetic transition was probed up to 45 T from the pulsed-field magnetization below 4.2 K \cite{High_field_parallel_b-2}. Since $a$ axis is the easy axis as suggested from the magnetic susceptibility \cite{PRL-93}, the phase diagrams for field along the $b$ and $c$ directions look very similar.

Temperature dependencies of $\kappa$ measured in various magnetic fields are displayed in Fig. \ref{kcT_Hc}. The $\kappa$ is strongly suppressed in 2 T and an approximate $T^{1.75}$ dependence is found at subKelvin temperatures, similar to $B \parallel b$. In higher fields, a broad peak is developed above 1 K, whereas the absence of the step-like decrease above the peak is different from the case of $B \parallel b$. In 14 T, a $T^{2.5}$ dependence is observed below 1 K, which reflects a weakened phonon scattering by spin fluctuations as compared with $B \parallel b$.

\begin{figure}
\includegraphics[clip,width=7.0cm]{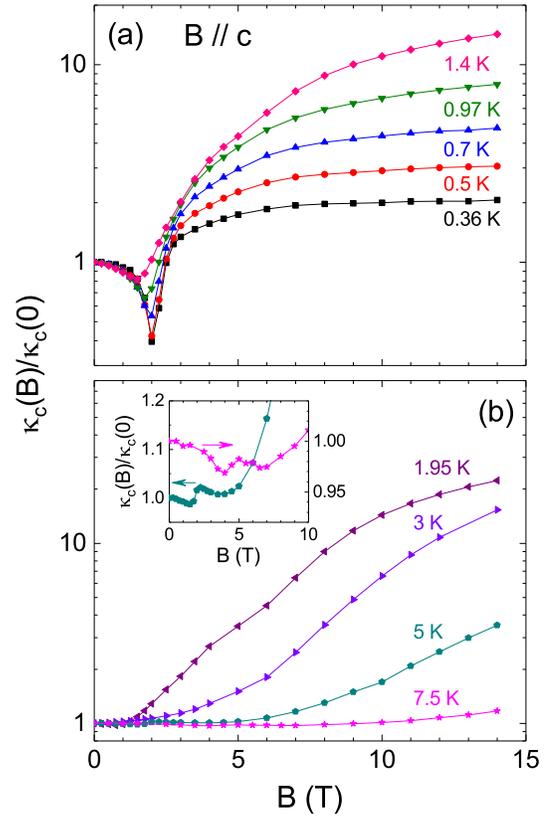}
\caption{(Color online) Magnetic field dependencies of $\kappa$ measured for $B \parallel c$ (a) below 1.4 K and (b) above 1.95 K. Inset to panel (b) is a zoom in of the data at 5 K and 7.5 K.}
\label{kcHc}
\end{figure}

The magnetic field dependencies of $\kappa$ in $B \parallel c$ in Fig. \ref{kcHc} resemble the situation for $B \parallel b$. At 0.36 K, a dip-like feature is present around 2 T with the amplitude reduced to 40\% of the zero-field value. Upon warming, the dip field is gradually decreased with a reduced amplitude. The feature is then smeared out at 3 K and the $\kappa$ is monotonously increased with the field. At higher temperatures, a double-valley behavior akin to the case of $B \parallel b$ is seen at 5 K, and it is shifted to higher fields at 7.5 K. In these two curves, the sharper dip at lower-field is likely associated to some magnetic transition, while the broader minimum at higher field should be related to the paramagnetic scattering. The ratio of $\kappa(B)/\kappa(0)$ in 14 T is also nonmonotonic with temperature. The largest $\kappa(B)/\kappa(0)$ is about 22 at 1.95 K, a little larger than that in $B \parallel b$ but still much smaller than that of Co$_3$V$_2$O$_8$.

As shown in Fig. \ref{Diagram_Hc}, the dip-like feature in $\kappa(B)$ below 2 K is related to the transition from C' to C phase. Above $T\rm_{LC}$, the weak anomaly in $\kappa(B)$ corresponds to the transition from LTI (or HTI) to C phase. Different from the phase diagram for $B \parallel b$, there is no similar high-field phase above LTI and HTI phases.

\section{Conclusions}

In this work, heat transport properties of the kagom\'{e}-staircase Ni$_3$V$_2$O$_8$ single crystals are studied, and the anisotropic phase diagrams are supplemented down to subKelvin temperatures. In zero field, the transition from C to C' phase, which was invisible in the magnetic susceptibility and neutron measurements, is detected from $\kappa(T)$. The minimum feature across both anomalies suggests that phonons are dominantly scattered by the critical spin fluctuations. When the magnetic field is applied along the $a$ axis, the $\kappa(B)$ data indicate an undetected magnetic phase below 1 K. In addition, $\kappa(B)$ is found to depend not only on the history but also on the magnitude of applying field. For $B \parallel b$, a high-field phase is detected above $T\rm_{LC}$, which is consistent with the early magneto-optical study.

Up to now, both the magnetism and magneto-electric coupling of Ni$_3$V$_2$O$_8$ has been well studied by means of various experimental techniques. Understanding the complicated magnetic properties is the basis for describing the mechanism and interesting multiferroic behaviors of this material. It is widely recognized that the rich phase diagrams are a competing result among the nearest-neighbor and next-nearest-neighbor exchange interactions, easy-axis anisotropy, anisotropic interaction, pseudodipolar interaction, and Dzyaloshinskii-Moriya interaction \cite{PRB-74}. However, in the previous works the study on the magnetism and magnetic transitions of Ni$_3$V$_2$O$_8$ at subKelvin temperatures is still missing from both the experimental and theoretical aspects, which prevents the community obtaining a comprehensive scenario of the ground state and a complete phase diagram. The discovery of a field-induced state below 1 K for $B \parallel a$ in this work provides important magnetism information at temperatures never achieved before. The nature of this state is not clear at present and will promote a further study on Ni$_3$V$_2$O$_8$. Experiments, such as magnetization, neutron diffraction, electrical polarization, etc. performed at ultra-low temperatures are in urgent need to uncover the physics of this newly discovered state.

\begin{acknowledgements}

This work was supported by the National Natural Science Foundation of China (Grant Nos. U1832209, 51702320, U1832166, 11574286, and 11874336), the National Basic Research Program of China (Grant Nos. 2015CB921201 and 2016YFA0300103) and the Innovative Program of Development Foundation of Hefei Center for Physical Science and Technology.

\end{acknowledgements}

\end{document}